\documentstyle[11pt,twoside]{article}
\begin{document}

\input epsf
 \noindent {\bf C. R. Acad. Sci. Paris, t.319  , S\'erie II , P.57-62 ,1994}\\
 \noindent Syst\`emes extra-galactiques/ {\it Extra-galactic Systems}
 \\
 \\
 \noindent {\Large {\bf Structure \`a grande \'echelle de l'Univers
 jusqu'\`a la distance de 200 Mpc.}}

 \begin{center}
 H\'el\`ene DI NELLA et Georges PATUREL
 \end{center}

 {\bf R\'esum\'e} --
 La distribution des galaxies jusqu'\`a une distance de 200 m\'egaparsecs
 (650 millions d'ann\'ee-lumi\`ere) est aplatie et montre une structure
 formant une coquille, approximativement centr\'ee sur le
 Super Amas Local (Amas Virgo). Ce r\'esultat confirme clairement
 l'existence de la structure hypergalactique mentionn\'ee en 1988.
 C'est actuellement la plus grosse structure jamais d\'etect\'ee.
 \\

 {\bf Large-scale structure of the Universe up to a distance of 200 Mpc.}
 \\

 {\it{\bf Abstract} -- The distribution of galaxies up to a distance of
 200 Mpc (650 million light-years) is flat and shows a structure like a shell
 roughly centered on the Local Supercluster (Virgo cluster). This result
clearly confirms the existence of the hypergalactic large scale structure
noted in 1988. This is presently the largest structure ever seen.}
\vspace{1.5cm}

{\it Abridged English Version} -- From projections of the whole sky it is
obvious that galaxies are not randomly distributed but that they are
merged in clusters and large structures of several tens of megaparsecs.
It is already known, after de Vaucouleurs studies ([8],[9]),
that nearby galaxies
form a flat structure, the Local Super cluster, the pole of which
is defined by galactic coordinates $l=47 \deg$; $b=6 \deg$.
Several studies ([5],[6],[7],[11]) tend to show
that distant clusters are located close to this plane.
More recently ([1], [3]) the existence of a flat large structure was
suspected with a pole at  $l=57 \deg$; $b=22\deg$.

{}From the Lyon-Meudon Extragalactic Database (LEDA), we built a
sample of 24324 galaxies with radial velocities smaller than $15000 km.s^{-1}$.
This sample will be used  to search
for the most populated plane. However, we have to take care
about the completeness of the sample to be sure that the result will not
come from a particular  feature of the sample itself.
The completeness was tested
form Rel.2 in which $N_{t}$ is the number of galaxies with diameter
greater than $D$.
$D$ is the apparent diameter at the brightness level
of $25 mag.arcsec^{-2}$ (D is expressed in 0.1' units). When the
sample is complete, the slope of Rel. 2 is constant, while it bends where
the sample begins to be incomplete. The completeness limit
for our sample is near $logD=1.2$. When the sample is restricted to
this limit (i.e. $logD>1.2$) the number of galaxies is only 5683. This
will constitute the complete sample used to search for the most populated
plane.

Thus, we counted galaxies within $\pm 15 deg$ around different planes
defined by their poles.  Pole directions were chosen over the range
$0 \deg$ to $90 \deg$ for the galactic latitude $b$ and  over the range
$0 \deg$ to $360 \deg$ for the galactic longitude $l$.
The results of these counts are given in Table 1, where it can
be seen that the position $lp=52 \deg; bp=16 \deg$defining the most populated
plane is quite stable with increasing distances.
In this zone, about 45\% of the complete sample lay in only 25\% of
the surface of the sphere. This pole will define the hypergalactic plane,
and will be used to define hypergalactic coordinates. The origin of
hypergalactic longitudes is arbitrarily defined as $l(origin)=l(pole)$
and $b(origin)=b(pole)-90\deg$.

The total sample of 24,324 galaxies with known radial velocity
is then used to make a representation of galaxy distribution within
a distance of $200 Mpc$.
The distance of each galaxy is
calculated trough the Hubble law (Rel. 1) assuming that
$H=75 km.s^{-1}.Mpc^{-1}$. The (X,Y,Z) coordinates were calculated from
Eq. 3--5, in which (hgl,hgb) are the hypergalactic longitude and latitude,
respectively. We projected on the X-Y hypergalactic plane all galaxies
with $|hgb|<15 \deg$. The result is given in Fig. 1. The most interesting
feature is the ring-like structure centered on the Local Super Cluster
(Virgo cluster).
A study in the X-Z plane suggests that the structure
is actually a shell (Fig. 2). The so-called {\it Big Wall} [2] is
a part of it. This shell is presently the largest structure ever seen.
Its kinematical and dynamical properties are not easy to understand.
Anyway, it confirms the existence of the hypergalactic structure
previously quoted [3].

\noindent ---------------------------------
 \section{Introduction}

 \noindent
 Les projections st\'er\'eographiques montrent que les galaxies
 ne sont pas distribu\'ees au hasard, mais qu'elles sont agglom\'er\'ees
 en groupes ou en amas
 de plusieurs centaines de galaxies s'\'etendant sur quelques
 dizaines de m\'egaparsecs.
 En 1953-1956, l'astronome G\'erard de Vaucouleurs a
 montr\'e que les galaxies proches (vitesse de r\'ecession
 plus petite que $3000 km.s^{-1}$) forment une structure plane connue
 maintenant comme le Super Amas Local. En coordonn\'ees galactiques le p\^ole
 de ce plan est $l=47 \deg$; $b=6 \deg$.
 En 1982 Zel'dovich, Einasto et Shandarin notent
 que les amas de galaxies lointains (vitesse sup\'erieure \`a 7000
$km.s^{-1}$),
 sont situ\'es pr\'ef\'erentiellement dans un plan. Ce r\'esultat est
confirm\'e
 par Shectman en 1985 puis par Tully (1986 et 1987) qui note que
 ce plan pr\'ef\'erentiel est voisin de celui du Super Amas Local.
 Au m\^eme moment, Bottinelli et al. (1986) signalent une structure
 aplatie visible sur une projection st\'er\'eographique de Flamsteed
(projection
 conservant les aires) r\'ealis\'ee avec des galaxies individuelles. Ils
donnent
 la position du p\^ole $l=57\deg$; $b=26\deg$, valeur r\'evis\'ee plus tard en
 $l=57\deg$; $b=22\deg$ (Paturel et al. 1988).

 \section{Recherche du plan privil\'egi\'e pour l'\'etude de la cin\'ematique}

 \noindent Dans l'objectif de se ramener \`a un plan dans lequel on
 puisse \'etudier plus facilement la cin\'ematique de l'Univers
 local, nous avons cherch\'e de mani\`ere syst\'ematique le plan
 sur lequel se situe le plus grand nombre de galaxies de l'Univers proche.
 Nous avons construit un \'echantillon de 24324 galaxies \`a partir de la
 base de donn\'ees extragalactiques des observatoires de Lyon-Meudon
 (LEDA), en extrayant pour chaque
 galaxie les param\`etres suivants: -- les coordonn\'ees galactiques
 $l$ et $b$ correspondant \`a la longitude et \`a
 la latitude respectivement -- la vitesse radiale de r\'ecession v
 qui sera utilis\'ee comme mesure de la distance avec la loi de Hubble.
 \begin{equation}
 d=v/H
 \end{equation}
 o\`u $d$ est la distance en m\'egaparsecs, $v$ la vitesse radiale
en $km.s^{-1}$ et $H$ la constante de Hubble
 en $km.s^{-1}.Mpc$. Dans tout ce travail la valeur de cette constante
 sera prise arbitrairement \'egale \`a 75 $km.s^{-1}.Mpc$ (c'est \`a
 dire une valeur moyenne entre les valeurs extr\^emes publi\'ees).

 \noindent Par ailleurs, afin que le r\'esultat ne d\'epende pas de
 l'\'echantillon
 utilis\'e, nous nous sommes assur\'es de la compl\'etude de l'\'echantillon,
 pour travailler avec toutes les galaxies
 de diam\`etre sup\'erieur \`a une certaine limite.
 Pour cela, nous avons construit la courbe

 \begin{equation}
logN_{t}=f(logD),
 \end{equation}

 o\`u $N_{t}$ est le nombre de galaxies de diam\`etre isophotal
 sup\'erieur \`a D.
 L'isophote adopt\'e pour la d\'efinition du diam\`etre est par convention
 de 25 magnitudes par seconde de degr\'e au carr\'e
 \footnote {L'isophote est tel que le flux par
  seconde de degr\'e au carr\'e correspondrait au flux d'une \'etoile
  de magnitude 25}
 et l'unit\'e de mesure
 est 0,1 minute de degr\'e.
 Pour les galaxies n'ayant pas de mesure de diam\`etre mais seulement une
 mesure de magnitude, nous avons d\'eduit le diam\`etre de la relation
       $logD = -0,2 m + 4$
 o\`u m est la magnitude apparente totale.

\noindent Il est facile de montrer que pour un univers
homog\`ene le nombre de galaxies augmente avec la distance selon une loi
en puissance (Paturel et al., 1994).
La pente de la
 relation 2 est alors constante tant que l'\'echantillon est complet.
 Au-del\`a, la courbe s'infl\'echit brusquement et le point de d\'ecrochage
 donne la limite de compl\'etude.
 Le r\'esultat de ce test est que la compl\'etude est bonne jusqu'\`a
 $logD=1,2$. Finalement, notre \'echantillon complet, contenant toutes
 les galaxies dont la vitesse de r\'ecession est inf\'erieure \`a
 $15000 km.s^{-1}$ et telles que $logD>1,2$, se restreint \`a
 5683 galaxies.

 \noindent
 Pour un plan d\'efini par les coordonn\'ees de son p\^{o}le: $lp$ et $bp$
 (longitude et latitude galactique respectivement), nous avons compt\'e
 le nombre de galaxies se trouvant dans une zone de $\pm 15\deg$ de part
 et d'autre du plan.
 Nous avons fait une recherche syst\'ematique du meilleur plan en donnant des
 valeurs de 0 \`a $90\deg$ \`a $b$ et de 0 \`a $360\deg$ \`a $l$.\\

 \noindent Nous pouvons voir dans la Table ~\ref{Tpole} le r\'esultat
 du comptage de galaxies:
 $d$ est la distance limite du comptage (donn\'ee en m\'egaparsec),
 $Ngal$ est le nombre total de galaxies situ\'ees dans cette limite,
 $lp$ et $bp$ sont les coordonn\'ees galactiques du p\^{o}le d\'efinissant
 le plan regroupant le plus grand nombre de galaxies et
 $\% gal$ est le pourcentage de galaxies se trouvant \`a $\pm 15\deg$ du
 plan pr\'ec\'edent.
 \noindent Si les galaxies \'etaient r\'eparties uniform\'ement dans la
 sph\`ere de comptage de rayon $d$, en regardant dans une zone de $30\deg$,
 c'est \`a dire d'angle solide $\pi$ st\'eradians, on devrait y trouver
 25\% du nombre total de galaxies de cette sph\`ere (proportion des
 angles solides).
 Dans la Table~\ref{Tpole}, on voit que le nombre de galaxies trouv\'ees
 dans le plan est bien sup\'erieur (44\%), ce qui montre que la r\'epartition
 des galaxies dans l'Univers local est plut\^ot aplatie.

\noindent Une position du p\^{o}le \`a $lp=52\deg$ et $bp=16\deg$ semble
 particuli\`erement stable entre 80 Mpc et 200 Mpc. Ce p\^ole est assez proche
 de celui trouv\'e en 1986 et 1988. Ce p\^ole d\'efinira
 le plan hypergalactique.
 D'apr\`es la Table~\ref{Tpole}, on peut \'egalement donner une nouvelle
 mesure de la position du p\^ole du Super Amas Local.
 En 1976 de Vaucouleurs donnait
 $lp=47\deg$ et $bp=6\deg$ pour les galaxies jusqu'\`a environ
 3000 $km.s^{-1}$, ici nous trouvons $lp=46\deg$ et $bp=14\deg$,
 pour la m\^eme r\'egion.

 \section{Repr\'esentation de l'Univers local dans le plan hypergalactique}

\begin{figure}
 \caption{
 Vue du plan hypergalactique. La trace de la coquille est visible.
 Sa position est approximativement centr\'ee sur l'amas Virgo.
 -- {\it Face-on view of the hypergalactic plane. The trace of
 the shell is visible. Its position is nearly centered on
 the Virgo cluster.}
 }
\epsfxsize=11cm
\epsfysize=8cm
\epsffile{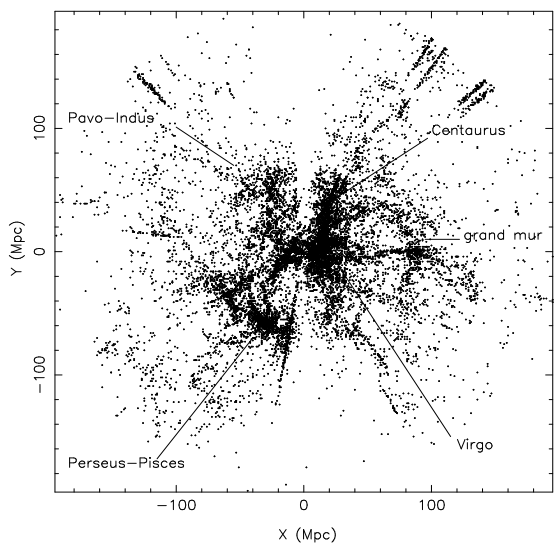}
 \label{planhyper}
 \end{figure}

 \noindent Par une transformation de coordonn\'ees, on peut repr\'esenter
 les galaxies de la zone $\pm 15\deg$ en projection
 sur le plan hypergalactique. La transformation est la suivante:
 \begin{equation}
        X= d.cos(hgb).cos(hgl)
 \end{equation}
 \begin{equation}
        Y= d.cos(hgb).sin(hgl)
 \end{equation}
 \begin{equation}
        Z= d.sin(hgb)
 \end{equation}
 Les coordonn\'ees hypergalactiques (hgl,hgb) sont calcul\'ees
 par un changement de
 rep\`ere \`a partir des coordonn\'ees galactiques du p\^ole et de l'origine
des
 longitudes.
 Cette origine des longitudes hypergalactiques est d\'efinie arbitrairement
 dans le plan hypergalactique aux coordonn\'ees:
 $l(origine)=l(pole)$ et $b(origine)=b(pole)-90\deg$.
 On peut voir cette repr\'esentation sur la figure Fig.~\ref{planhyper}.

\begin{figure}
 \caption{
 Vue perpendiculaire au plan hypergalactique. Une trace similaire
 \`a celle que nous avions dans le plan hypergalactique est encore visible,
 ce qui fait penser que la structure \`a la forme d'une coquille
 ellipso\"{\i}dale.
 --{\it View perpendicular to the hypergalactic plane. A trace
similar to the one seen on the hypergalactic plane is still visible.
This suggests that the structure has the shape of an ellipsoidal shell.}
}
\epsfxsize=11cm
\epsfysize=8cm
\epsffile{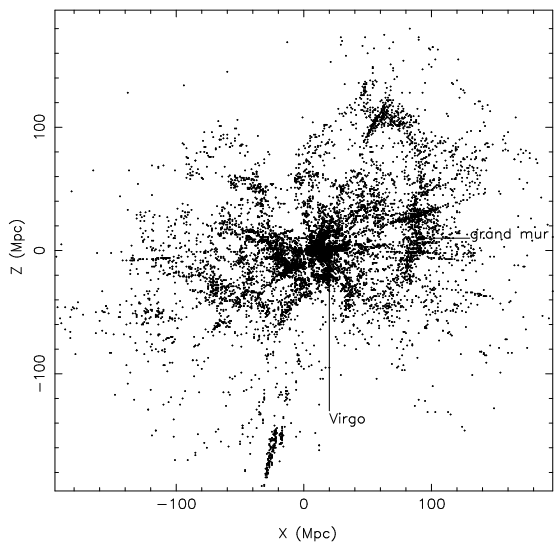}
 \label{planperpan}
 \end{figure}

 Cette repr\'esentation appelle plusieurs remarques:
\begin{itemize}
\item Des structures radiales tr\`es allong\'ees sont visibles, par exemple \`a
 la p\'eriph\'erie du graphique.
 Ces structures r\'esultent de l'incertitude sur la distance
 d\'eduite de la vitesse radiale. En effet la vitesse radiale
 mesur\'ee n'est pas seulement la vitesse cosmologique (fonction de la
 distance). Elle compte aussi la composante radiale de la dispersion
al\'eatoire
 des vitesses. Il appara\^{\i}t donc une dispersion sur les distances, alors
 que pratiquement aucune erreur n'existe en direction. Les amas de galaxies
 apparaissent donc \'etir\'es le long de la ligne de vis\'ee. La cons\'equence
 pratique est qu'il ne faudra pas accorder trop de r\'ealit\'e aux structures
 radiales qui peuvent provenir de cet artefact.

\item On voit \'egalement une zone (verticale sur le graphique) o\`u aucune
 galaxie n'appara\^{\i}t. Cette zone r\'esulte de l'absorption due \`a la
 poussi\`ere du disque de notre Galaxie qui cache les autres galaxies de
 l'Univers dans cette direction. Les relev\'es dans l'infrarouge devraient
 permettre de combler partiellement cette zone d'absence.

\item Le r\'esultat le plus int\'eressant visible sur la Fig.~\ref{planhyper}
 est la structure en forme d'anneau centr\'ee sur le Super Amas Local
(amas Virgo) et d'un rayon d'environ
 90 Mpc (pour $H=75 km.s^{-1}.Mpc^{-1}$).
On ne peut pas invoquer l'artefact vu pr\'ecedemment puisque
 cette structure n'est pas radiale.
 Une projection perpendiculaire au plan hypergalactique (Fig.~\ref{planperpan})
 permet de dire que cet anneau est la trace d'une
 coquille ellipso\"{\i}dale, sensiblement centr\'ee sur le Super amas local.
 La structure connue sous le nom de "grand mur" (Huchra et al., 1990)
 appara\^{\i}t \^{e}tre une portion de cette coquille.
\end{itemize}

 \section{Conclusion}
 \noindent Avec un \'echantillon complet en diam\`etre apparent, nous avons
 trouv\'e un plan privil\'egi\'e dans l'Univers local. Ce plan est
 identifi\'e au plan hypergalactique trouv\'e avec la base de donn\'ees
 de Lyon-Meudon  (Bottinelli et al. 1986; Paturel et al. 1988).
 Ce plan s'identifie \`a plus grande \'echelle avec le plan trouv\'e par
 Tully (1986).
 La repr\'esentation des galaxies dans ce plan fait
 appara\^{\i}tre la trace d'une coquille
 ellipso\"{\i}dale aplatie approximativement centr\'ee sur le Super Amas Local
 (c'est \`a dire approximativement sur le centre de l'amas Virgo).
 Ce r\'esultat confirme clairement l'existence de la tr\`es grande
 structure {\it hypergalactique} mentionn\'ee par Paturel et al. (1988).

 Cette structure est int\'eressante  car c'est actuellement la
plus grande structure connue.
Son \'etude tant cin\'ematique que dynamique pose quelques probl\`emes.
Si la gravitation est responsable de la forme \`a sym\'etrie centrale,
on  comprend mal comment l'\'equilibre d'une telle structure pourrait
\^etre assur\'e.

\vspace{1.5cm}

\small
 REFERENCES BIBLIOGRAPHIQUES \\

   [1]     L. BOTTINELLI, P. FOUQUE, L. GOUGUENHEIM, G. PATUREL,
 {\it Structure \`a grande \'echelle visible avec la base de donn\'ees
 extragalactiques}, 1986, La dynamique des structures gravitationnelles.

   [2]    J.P. HUCHRA, J.P. HENRY, M. POSTMAN et M.J. GELLER,
 1990, The Astrophysical Journal, 365, 66

   [3]   G. PATUREL, L. BOTTINELLI, L. GOUGUENHEIM et P. FOUQUE,
 {\it New determination of the pole of a "hypergalactic" large-scale
 system}, Astronomy and Astrophysics, 1988, 189, 1

   [4]   G. PATUREL, H. DI NELLA, L. BOTTINELLI, P. FOUQUE et L. GOUGUENHEIM
 {\it Interpretation of a completeness curve}, Astronomy and Astrophysics,
 1994 (Sous-presse)

   [5]   S.A. SHECTMAN, 1985, The Astrophysical Journal Supplement Series,
 57, 77

   [6]   R.B. TULLY,
 {\it Alignment of clusters and galaxies on scales up to 0.1c},
  1986, The Astrophysical Journal, 303, 25

   [7]   R.B. TULLY,
 {\it More about clustering on a scale of 0.1c},
 1987, The Astrophysical Journal, 323, 1

   [8]   G. DE VAUCOULEURS, 1953, Astronomical Journal, 58, 30

   [9]   G. DE VAUCOULEURS, 1956, Vistas Astr., 2, 1584

   [10]   G. DE VAUCOULEURS, A. DE VAUCOULEURS, JR. H.G. CORWIN,
 {\it Second Reference Catalogue of Bright Galaxies}, 1976, University of Texas
 Press, Austin

   [11]   YA. B. ZEL'DOVICH, J. EINASTO et S.F. SHANDARIN, 1982, Nature, 300,
407

\vspace{1cm}

{\it
\hspace {8cm}  \noindent Observatoire de Lyon

\hspace {8cm}  \noindent 69561 Saint-Genis Laval CEDEX, France

\hspace {8cm} \noindent t\'el\'ephone: 78.86.83.83

\hspace {8cm} \noindent t\'el\'ecopie: 78.86.83.86

\hspace {8cm} \noindent e-mail: patu@adel.univ-lyon1.fr
}

\newpage
 \begin{table}
 \caption{
Nombre de galaxies situ\'ees \`a une distance inf\'erieure \`a d et
pourcentage (\%gal) de galaxies localis\'ees
 \`a $\pm 15\deg$ du plan d\'efini par le p\^ole de coordonn\'ees
 $lp$ et $bp$.
Ce pourcentage est tr\`es sup\'erieur \`a ce que laisse attendre
une distribution homog\`ene des galaxies (voir le texte).
 -- {\it Number of galaxies located at a distance smaller than d and
 percentage (\%gal) of galaxies located
 at $\pm 15\deg$ of the plane defined by the pole of coordinates
$lp$ and $bp$. This percentage is much higher than the one expected
for an homogeneous distribution of galaxies (see text).}
}
\vspace{1cm}
\label{Tpole}
\begin{tabular}{rllllllllll}
\hline
 $d (Mpc)$   : &  20 &  40 &  60 &  80 & 100 & 120 & 140 & 160 & 180 & 200 \\
 $Ngal$        : & 1037& 2774& 3945& 4939& 5310& 5507& 5593& 5643& 5661& 5683\\
 $lp (\deg)$ : &  46 &  46 &  46 &  51 &  52 &  52 &  52 &  52 &  52 &  52 \\
 $bp (\deg)$ : &  14 &  14 &  17 &  16 &  16 &  16 &  16 &  16 &  16 &  16 \\
 $\% gal $     : &  55 &  46 &  46 &  45 &  45 &  44 &  44 &  44 &  44 &  44 \\
\hline
\end{tabular}
\end{table}

\newpage

 \end{document}